\documentclass[a4paper]{article}

\usepackage{INTERSPEECH2019}
\usepackage{multirow}
\usepackage{makecell}
{

\title{Deep Neural Baselines for Computational Paralinguistics}
\name{
Daniel Elsner$^{1,2}$, 
Stefan Langer$^1$,
Fabian Ritz$^1$,
Robert Mueller$^1$,
Steffen Illium$^1$
}
\address{
  $^1$LMU Munich, Germany\\
  $^2$Tawny GmbH, Germany}
\email{
\{extern.daniel.elsner,stefan.langer,fabian.ritz,robert.mueller,steffen.illium\}@ifi.lmu.de
}

\begin{document}

\maketitle
\begin{abstract}
Detecting sleepiness from spoken language is an ambitious task, which is addressed by the Interspeech 2019 Computational Paralinguistics Challenge (ComParE).
We propose an end-to-end deep learning approach to detect and classify patterns reflecting sleepiness in the human voice.
Our approach is based solely on a moderately complex deep neural network architecture.
It may be applied directly on the audio data without requiring any specific feature engineering, thus remaining transferable to other audio classification tasks.
Nevertheless, our approach performs similar to state-of-the-art machine learning models.
\end{abstract}
\noindent\textbf{Index Terms}: affective computing, speech recognition, deep learning, computational paralinguistics, ComParE
\section{Introduction}
\label{sec:intro}
\emph{Computational paralinguistics} describe a research field that has been established in the past twenty years through remarkable research progress, e.g. in the area of automated affect recognition or determining illnesses from speech~\cite{schuller2013computational,grishman1986computational,kiss2016language}.
More generally, emphasis is put on the analysis of the human voice to design~\emph{affective computers} that possess empathic competencies such as recognizing, expressing, modeling, or communicating emotions~\cite{picard2003affective}.
In most paralinguistic problems, task-specific feature engineering and model tuning allows to build state-of-the-art statistical learning models, e.g. machine learning (ML) classifiers~\cite{schuller2013paralinguistics}.
Powerful software packages such as the open-source toolkits \emph{OpenSMILE}~\cite{eyben2010opensmile}, \emph{OpenXBOW}~\cite{schmitt2017openxbow} and \emph{AuDeep}~\cite{freitag2017audeep} were created to extract these relevant features from raw audio data.
However, more general approaches ease the transfer of models to familiar problem domains without prior expert knowledge about audio signal processing, while still providing adequate predictive performance.
Recent work in deep learning focuses on end-to-end modeling of affect recognition problems, i.e. training models on raw signal data without extensive data pre-processing~\cite{graves2014towards}.
Even though deep learning has revolutionized artificial intelligence (AI) research areas, e.g. computer vision and machine translation, the full potential of Deep Neural Networks (DNNs) can usually only be utilized with datasets of sufficient quality and quantity~\cite{rouast2019deep}.
The availability of public databases for affect recognition problems has especially improved through annual competitions and challenges.\\
This paper proposes an end-to-end deep learning approach for the Interspeech 2019 Computational Paralinguistics Challenge (ComParE)~\cite{schullerinterspeech}.
While avoiding task-specific feature engineering and providing an agnostic approach for modeling paralinguistic problems, it addresses the aforementioned growing demand for general approaches in computational paralinguistics and affective computing.
Our main contribution is a moderately complex depp neural network (DNN) architecture that is able to detect and classify sleepiness in human voice from the SLEEP corpus.
We show that its performance is comparable to the provided baseline which fuses three support vector machines (SVMs) trained on more than 6,000 extracted audio features from the toolkits \emph{OpenSMILE}, \emph{OpenXBOW} and \emph{AuDeep}.
Also, we perform a sanity check on its transferability.
\section{Deep Learning in Computational Paralinguistics}
\label{sec:deep-learning-in-paralinguistics}
Throughout recent years there has been a shift in affective computing from classical ML approaches towards deep learning~\cite{rouast2019deep}.
Among other areas, this also concerned computational paralinguistics.
The inherent absence of task-specific, manual feature extraction and selection in deep learning allows researchers to design complex, non-linear models.
Those either reveal useful (latent) feature embeddings (e.g., unsupervised learning with deep auto encoders~\cite{freitag2017audeep}) or learn feature representations directly from unstructured data for predictive modeling (e.g., classification of images with CNNs~\cite{krizhevsky2012imagenet}).
Generally, DNNs allow end-to-end problem modeling, where raw data (e.g. audio signal data) is fed to the model and a prediction, e.g. a class label or a continuous value, is returned.
To summarize, three major causes have enabled this shift, namely (i) increased computing capabilities and (ii) learning capacity and precision of deep learning models through wider and deeper neural networks, as well as (iii) an ever growing amount of free-to-use, labeled datasets publicly available~\cite{rouast2019deep}.\\
DNNs can be considered the state-of-the-art technology to human affect recognition software~\cite{cummins2018speech}.
To tackle the problem of diverse data representations, DNNs can be altered accordingly in order to enhance their performance.
Within the field of computational paralinguistics CNNs, as well as Recurrent Neural Networks (RNNs), are amongst the best known network architectures. 
This work emphasizes the former, which are foremost established and applied numerously~\cite{krizhevsky2012imagenet,liang2015recurrent,lawrence1997face} in the field of image processing.\\
Moreover, CNNs are able to learn features representations of different abstraction levels, ranging from low level features, such as edges, to high level features, such as eyes or noses, natively~\cite{rouast2019deep,cummins2018speech}.
Those deep spatial features outperform handcrafted spatial features in most cases~\cite{rouast2019deep,cummins2018speech}.
Another major advantage of CNNs is the low amount of parameters, compared to other network types like RNNs, and hence faster computing times during model training.
CNNs can be applied in multiple dimensionalities, such as 3D, mostly used for video analysis, 2D for the aforementioned image processing tasks, as well as 1D for one dimensional raw data inputs.
Therefore, CNNs have been an attractive option to researchers in computational paralinguistics, starting to be applied as early as 2011~\cite{jaitly2011learning}.
\section{Related Work}
\label{sec:related-work}
Greeley et al. investigate the effect of sleepiness on war fighters and civilian pilots and how machines could non-intrusively detect fatigue in very noisy environments~\cite{greeley2006detecting}.
The authors report that changes in single discrete voice parameters are not sufficient to detect whether the speaker suffers from fatigue.
Therefore, they propose a more holistic approach, combining the coefficients of the cepstral transformation with an automatic speech recognition system.
This correlation-based voice metric achieved on-par results with state-of-the-art fatigue measurements.
Krajewski et al. follow a somewhat classical ML procedure towards measuring fatigue in audio recordings~\cite{krajewski2009acoustic}.
Their results are based on speech characteristics such as prosody, articulation, and speech-quality-related and reach a classification accuracy of 86.1\% on data collected in a sleep deprivation study.
Their top result was achieved utilizing an SVM.
Within the ComparE challenge 2010, Marie-José Caraty and Claide Montacié address the problem of detecting vocal fatigue based on the statement that it highly affects the work in some professions~\cite{caraty2014vocal}.
The authors build their investigations upon three experiments, being prosodic analysis, a two-class SVM classifier, and a combination of multiple phoneme-based comparison functions.
The two-class SVM model reached an unweighted accuracy of 68.2\%.
The authors state that this suggests the feasibility of vocal fatigue detection.
Recently, Cummins et al. revisited approaches and results of past years of paralinguistics challenges and outline a noticeable shift from classical ML techniques towards deep learning models~\cite{cummins2018speech}.
Accordingly, the majority of submissions made use of some sort of deep learning approach.
Participants utilized DNNs for feature representation learning, classification or the combination of both, which underlines the growing interest in DNNs.
In 2011 there has been a ComParE sub-challenge, asking participants to distinguish between \emph{sleepy} and \emph{not-sleepy}.
The best solution at the time can be considered a classical ML solution, and has not been challenged through an end-to-end deep learning approach as of 2017.
This leads to the assumption that there is a gap in research approaches concerning sleepiness detection and investigating more general, non task-specific approaches is reasonable.
\section{Paralinguistic Problem Modeling}
\label{sec:problem}
The datasets provided for the 2019 ComParE challenges -- in their original form -- are not suited specifically for deep learning due to the limited number of samples.
Thus, we experimented with pre-processing as well as augmentation techniques and tested different CNN architectures.
These steps are outlined in the following subsections and evaluated in section~\ref{sec:results}.
\subsection{SLEEP Corpus}
\label{sec:problem:sub:sleep}
\begin{figure}
	\includegraphics[width=\columnwidth]{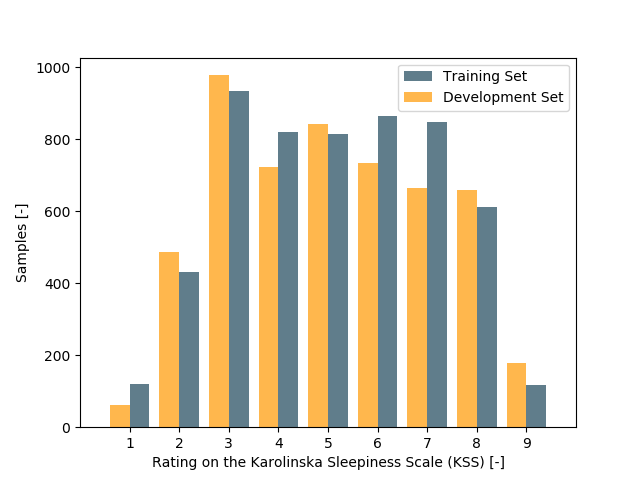}
	\caption{KSS rating distribution of the training and development set.}
	\label{fig:class_distribution}
	\vspace{-3mm}
\end{figure}

The SLEEP corpus consists of $16,462$ audio samples, each with a duration of about four seconds, which are split into \emph{training}, \emph{development} and \emph{test} set.
Its audio files have sampling rates of 16 kHz with a quantization of 16 bit.
The audio samples were gathered from $915$ subjects performing different speaking tasks such as reading out given text passages.
Recordings were carried out between 6 p.m. and midnight.
Afterwards, participants and post-hoc observers had to report sleepiness on the Karolinska Sleepiness Scale (KSS), ranging from 1 (extremely alert) to 9 (very sleepy).
These ratings were averaged to build the final label per recording.
The dataset was created at the Institute of Psychophysiology, Duesseldorf, Germany and the Institute of Safety Technology, University of Wuppertal, Germany.
More detailed statistics about the dataset and the label distribution are provided in Table~\ref{table:dataset_stats} and Figure~\ref{fig:class_distribution}, respectively.
The task of the challenge is to build a regression model that is able to predict the KSS rating for an audio recording~\cite{schullerinterspeech}.
\begin{table}[th]
\caption{Number of samples for training, development and test set as well as the respective mean ($\mu$), standard deviation ($\sigma$), minimum and maximum duration of the samples (in seconds).}
\label{table:dataset_stats}
\begin{tabular}{l|l|l|l|l|l}
Dataset     & Samples    & $\mu$ & $\sigma$ & min. & max.  \\ \hline
Training    & 5,564      & 3.87 & 0.64    & 1.56 & 5.00 \\
Development & 5,328      & 3.87 & 0.65    & 1.57 & 5.00 \\
Test        & 5,570      & 3.86 & 0.63    & 1.59 & 5.00 \\
\end{tabular}
\end{table}
\subsection{Data Pre-processing and Augmentation}
\label{sec:problem:sub:data}
In classical ML problem modeling, feature engineering and feature selection are integral parts of the data preparation process.
Hereby, the available data is reduced to relevant pieces of information, e.g. by excluding irrelevant data or modeling specific features.
Our approach omits the extraction of problem-specific features, but relies on the DNN models being capable of learning relevant feature representations themselves.
Nevertheless it is necessary to synthetically increase the training data volume to model the regression task as a deep learning problem.
The following subsections delineate pre-processing steps and augmentation techniques.
\subsubsection{Audio Pre-processing}
\label{sec:problem:sub:data:subsub:audio}
We experimented with different down-sampling rates for the provided audio files.
This is inspired by narrowband telephony only transmitting frequencies up to 4 kHz but still retaining the majority of information.
With respect to the Nyquist-Shannon sampling theorem, we choose a down-sampling to at least 8 kHz.
As a side effect, the down-sampling reduces the amount of data thus increases processing speed.
\subsubsection{Sliding Windows}
\label{sec:problem:sub:data:subsub:window}
As mentioned previously, the small volume of training data ($5,564$ samples) needs to be increased by magnitudes to suit a deep learning set up.
Therefore, we slice windows from a single audio recording in a sliding window manner.
We experimented with varying window sizes, as well as different strides, i.e. step sizes.
Consequently, we do not obtain one data point per audio sample, but get $(L-w) / s$ overlapping windows for a sample of length $L$ with window size $w$ and stride $s$.
The best results were achieved with $w=1.5s$, and $s=100ms$, that, given an exemplary sample of length $L=4s$, results in $(4s-1.5s) / 100ms = 25$ samples of the same label.
The total amount of data samples extracted from the training dataset is $134,395$ and $128,808$ from the development dataset.
\subsubsection{Data Up- and Downsampling}
\label{sec:problem:sub:data:subsub:upsampling}
As described in section~\ref{sec:problem:sub:data}, the distribution of samples is imbalanced at the extrema of the KSS.
To prevent the models from overfitting on the large corpus of samples labelled between 3 and 8, we performed up- and down-sampling of under- and overrepresented samples, respectively.
Consequently, the extracted sample windows labelled 1 or 9 (extremely alert and very sleepy) were included multiple times into the dataset, i.e. up-sampled, whereas sample windows labelled between 3 and 8 were only partly included, i.e. down-sampled.
\subsubsection{Data Augmentation}
\label{sec:problem:sub:data:subsub:augmentation}
To further increase the volume of available training data, we tested the following augmentation techniques:\\
\textbf{Reversing samples:} We flipped each sample window and included both the reversed and the  original sample in the training set.
The hypothesis was that the relevant patterns might be independent of the exact sequential structure of the sample.\\
\textbf{Background overlay:} To make the models more robust, we overlayed the training samples with background noise manually extracted from parts of training recordings where voice was not present.
Again, the original as well as the modified samples were included for training.
This is a common approach and not limited to the available training data since background noise can be recorded separately for different background settings.\\
\textbf{Noisy labelling}: With previously mentioned sliding window approach, all windows extracted from a single sample were labelled with the same KSS rating.
As the actual self-reported degree of sleepiness may not be occurrent during the whole sample, we applied noise randomly taken from a normal distribution onto the KSS ratings.
The hypothesis was that this bias might lead to more robust models, as the KSS ratings could have been ambiguous at the time of labelling (i.e. participants' self-reports compared to the post-hoc observers').
We included both the samples with the actual and those with noisy labels in the training dataset.
\subsection{Model Architecture}
\label{sec:problem:sub:model}
The following describes our DNN architecture and the mechanism necessary both aggregate the predicted labels of the sliding windows and map them back onto the original audio samples.
\subsubsection{Prediction Aggregation Mechanism}
\label{sec:problem:sub:model:subsub:pred}
In section~\ref{sec:problem:sub:data:subsub:window} we outlined the sliding window approach to increase the volume of the dataset.
As this approach leaves us with $n$ samples per audio file (depending on the file's length $L$, window size $w$, and stride $s$), a mechanism for merging the predicted labels (i.e. KSS ratings) during the inference to finally generate one label for the original audio sample is needed.
For example, if an audio file $a$ leads to $25$ windows, which were fed to the model during inference (i.e. prediction), $25$ predicted labels, one for each window, have to be merged and mapped back onto the original audio sample $a$.
We aggregated the predicted labels in two ways, by taking the mean and the median of the predictions.
Ultimately, we clipped the resulting prediction for the original audio sample into the KSS rating from 1 to 9 and performed a typecast to an integer number.
\subsubsection{Convolutional Neural Network}
\label{sec:problem:sub:model:subsub:cnn}
\begin{figure}
	\includegraphics[width=\columnwidth]{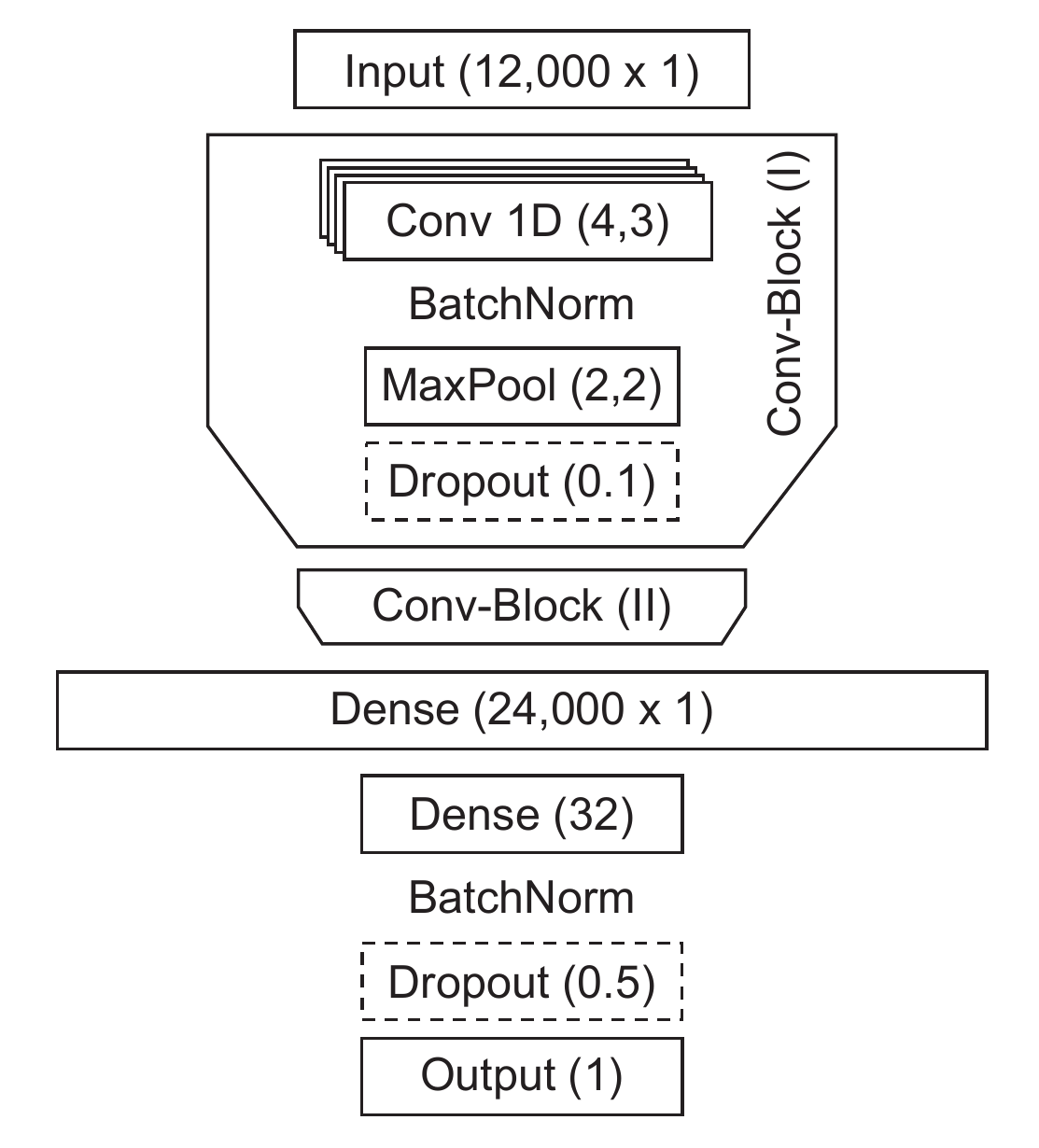}
	\caption{Proposed DNN Architecture}
	\label{fig:dnn_architecture}
	\vspace{-3mm}
\end{figure}

Figure~\ref{fig:dnn_architecture} depicts the proposed DNN architecture that is capable of learning spatial -- thus short sequential -- feature representations from raw audio data.
The DNN consists of two 1D convolutional layers, each with four filters and kernel size tree, that are connected through batch normalization and max pooling layers, each having the size of the vector.
After these convolutional blocks (Conv-Blocks), one fully connected layer with 32 neurons leads to a final dense layer with a single linear activation unit.
The regression model was trained using the mean squared error (MSE) loss and the Adam optimization algorithm with a learning rate of $0.001$ without decay.
Except for the final dense layer, ReLU activations were used.
To prevent overfitting during training, a dropout was applied after the convolutional (drop rate 0.1) and dense layers (drop rate 0.5)
Evaluated on the development dataset, the described model resulted in the best model.
However, we discuss further (hyper-) parameters in section~\ref{sec:results:sub:devel}.
\section{Results}
\label{sec:results}
The baselines' results are reported with Spearman’s rank correlation coefficient $ \rho $~\cite{schullerinterspeech}.
Since the actual labels of the test set are not available ex-ante, we trained our models on the training set and tested on the development set.
We then submitted the best model to the challenge and reported the score on the test set.
For the sake of interpretation, we include our loss metric MSE as well as the mean absolute error (MAE).
\subsection{Development Dataset}
\label{sec:results:sub:devel}
\begin{figure}
	\includegraphics[width=\columnwidth]{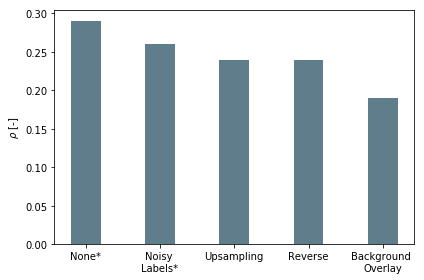}
	\caption{Overview of Spearman's rank correlation coefficient ($ \rho $) of applied pre-processing and augmentation techniques compared to the best model without further processing (\textbf{None}) on the development set.
	Models denoted with * were submitted to the challenge.}
	\label{fig:augmentation}
	\vspace{-3mm}
\end{figure}

\begin{table}[th]
\vspace{-2mm}
\caption{Evaluation scores for MSE, MAE, and $\rho$ on the development set with different parameter combinations regarding (i) sample rate, (ii) window size, (iii) amount of convolutional blocks.
The proposed approach is varied by each one of the three parameters while the others remain fixed.
Note that the stride was fixed at $0.1 s$.
Models denoted with * were submitted to the challenge.
}
\label{tab:dev_results}
\centering
\begin{tabular}{l|r|r|r}
                                                    & MSE  & MAE  &$\rho$\\ \hline
\multicolumn{1}{l|}{Strongest Baseline}             & --   & --   & 0.27 \\ \hline
\multicolumn{1}{l|}{\makecell[l]{{Proposed Approach*}\\
(16 kHz, 1.5 s, 2 Conv-Blocks)}}                    & 4.44 & 1.72 & 0.29 \\ \hline
\multicolumn{1}{l|}{Smaller Window Size (1 s)}      & 3.96 & 1.67 & 0.28 \\
\multicolumn{1}{l|}{Smaller Sampling Rate (8 kHz)*}  & 3.89 & 1.65 & 0.26 \\
\multicolumn{1}{l|}{More Conv-Blocks (3)}           & 4.12 & 1.70 & 0.24 \\
\end{tabular}
\end{table}
To allow proper comparison of models with (hyper-) parameters, we set the batch size to $64$ and the amount of epochs to $8$ for training.
This allowed testing multiple parameter combinations without large computational and time-consuming overhead.
Table~\ref{tab:dev_results} shows the evaluation scores for MSE, MAE, and $\rho$ on the development set for different parameters and models. 
The results imply that our models perform slightly better ($\rho \approx 0.29 \pm 0.03$) than the strongest baseline models ($\rho \approx 0.26$) on the development set.
Generally, we found that during training, the loss was monotonically decreasing on the training and the development set.
However, most models tend to overfit on the training set after $6$ epochs and we therefore saved the best models during the entire training process based on their loss on the development set.
The different pre-processing and augmentation techniques described in section~\ref{sec:problem:sub:data:subsub:augmentation} did not produce better results regarding $\rho$ (see figure~\ref{fig:augmentation}).
However, a combination of techniques is beyond the scope of this work as we aim to provide a generic approach rather than a tailored solution for this specific problem.
\subsection{Test Dataset}
\label{sec:results:sub:test}
After experimenting with different model (hyper-) parameters, we selected the best performing model, trained it again on the entire development set, and submitted the resulting model to the ComparE challenge.
As we only received $\rho$, MSE and MAE cannot be reported.
Note that the distribution of predicted labels is very similar to the actual distribution of labels in the development and training set (underrepresentation of 1, 2, 8, and 9).
The best provided baseline, an ensemble of three SVMs, achieved $\rho=0.343$.
To test different parameter configurations and augmentation techniques, three of the described models were submitted to the challenge and performed as follows:
\begin{enumerate}
	\item The proposed approach with smaller sampling rate (8 kHz) achieved $\rho=0.28$ on the test set.
	\item The proposed approach with noisy labels (see figure~\ref{fig:augmentation}) achieved $\rho=0.302$ on the test set.
	\item The proposed approach with default sampling rate (16 kHz) and without augmentation achieved $\rho=0.335$ on the test set.
\end{enumerate}
The two left submissions out of five possible submissions were reserved for experimentation with other approaches within the challenge.
\subsection{Transferability}
\label{sec:results:sub:transferability}
In order to evaluate the transferability of our DNN architecture, a similar model was trained to classify Styrian Dialects within the respective ComParE sub-challenge.
However, instead of solving a regression problem, the final single linear activation unit was replaced with a three-unit softmax activated dense layer.
On the development set, our best model achieved an Unweighted Average Recall (UAR) of $44.00\%$.
The same model scored an UAR of $38.28\%$ on the test set in our submission to the challenge.
Compared the three strongest baseline models with an average of $40.00\%$ on the test set, we claim that transferring our DNN architecture to other audio classification tasks is feasible.
\section{Conclusion}
\label{sec:conclusion}
This paper presented an end-to-end deep learning approach to detect and classify sleepiness in the human voice.
The proposed 1D convolutional DNN is capable of learning spatio-temporal feature representations from raw audio data.
Its performance is comparable with models trained on current audio feature extraction toolkits.
Moreover, we performed a sanity check on the transferability of our approach.
The results indicate that our DNN architecture may be used as a problem-agnostic and straightforward baseline in addition to classical ML approaches.
The end-to-end method emphasizes generalizability and transferability to other domains, e.g. in computational paralinguistics, contrary to problem-specific feature engineering.
Our proposed architecture is especially suitable for context aware multimedia recommendation systems.
In a possible use-case, the system could recommend e.g. radio stations or songs dependent on the fatigue level, detected in the user's voice.
Future work could deepen the extent of architecture search and parameter tuning as other automatic ML approaches suggest, to ultimately further democratize AI research~\cite{zoph2018learning}.

\paragraph*{Acknowledgements.}
The HRADIO project and thus this work was funded by H2020, the EU Framework Programme for Research and Innovation.

\bibliographystyle{IEEEtran}

\newpage
\bibliography{mybib}

\end{document}